\documentclass[]{aa}

\usepackage{amsmath}
\usepackage{graphicx}
\usepackage{natbib}
\usepackage{placeins}
\usepackage[flushleft]{threeparttable}
\usepackage{txfonts}
\usepackage{upgreek}
\bibpunct{(}{)}{;}{a}{}{,}

\begin{document}

\title{Properties of a hypothetical cold pulsar wind in LS~5039}

\author{V. Bosch-Ramon\inst{1}}
  
\institute{Departament de F\'{i}sica Qu\`antica i Astrof\'{i}sica, Institut de Ci\`encies del Cosmos (ICC), Universitat de Barcelona (IEEC-UB), Mart\'{i} i Franqu\`es 1, E08028 Barcelona, Spain. \\\\  \email{vbosch@fqa.ub.edu}}

%\offprints{V. Bosch-Ramon, \email{vbosch@fqa.ub.es}}

\titlerunning{On the hypothetical pulsar wind of LS~5039}

\abstract{LS~5039 is a powerful high-mass gamma-ray binary that probably hosts a young non-accreting pulsar. However, despite the wealth of data available, the means by which the non-thermal emitter is powered are still unknown.} 
{We use a dynamical-radiative numerical model, and multiwavelength data, to constrain the properties of a hypothetical pulsar wind that would power the non-thermal emitter in LS~5039.} 
{We ran simulations of an ultrarelativistic (weakly magnetized) cold $e^\pm$-wind that Compton scatters stellar photons and that dynamically interacts with the stellar wind. The effects of energy losses on the unshocked $e^\pm$-wind dynamics, and the geometry of the two-wind contact discontinuity, are computed for different wind models. The predicted unshocked $e^\pm$-wind radiation at periastron, when expected to be the highest, is compared to LS~5039 data.} 
{The minimum possible radiation from an isotropic cold $e^\pm$-wind overpredicts the X-ray to gamma-ray fluxes at periastron by a factor of $\sim 3$. In the anisotropic (axisymmetric) wind case X-ray and $\gtrsim 100$~MeV data are not violated by wind radiation if  the wind axis is at $\lesssim 20-40^\circ$ from the line of sight (chance probability of $\lesssim 6-24$\%), depending on the anisotropic wind model, or if  the wind Lorentz factor $\in 10^2-10^3$, in which case the wind power can be higher, but it requires $e^\pm$-multiplicities of $\sim 10^6$ and $10^9$ for a $10^{-2}$~s and  10~s pulsar period, respectively.} 
{The studied model predicts that a weakly magnetized cold pulsar $e^\pm$-wind in LS~5039 should be strongly anisotropic, with either a wind Lorentz factor $\in 10^2-10^3$ and very high multiplicities or with a fine-tuned wind orientation. A weakly magnetized, cold baryon-dominated wind would be a possible alternative, but then the multiplicities should be rather low, while the baryon-to-$e^\pm$ energy transfer should be very efficient at wind termination. A strongly magnetized cold wind seems to be the most favorable option as it is consistent with recent research on pulsar winds and does not require fine-tuning of the pulsar wind orientation, and the wind multiplicity and Lorentz factor are less constrained.}


\maketitle

\section{Introduction}\label{intro}

LS~5039 is a strong galactic source of variable and periodic gamma rays
\citep{par00,aha05,fer09,had12,col14,cha16}. This source displays non-thermal radio emission with
milliarcsecond to subarcsecond components \citep{par00,par02} and X-rays of likely non-thermal origin
\citep[see, e.g.,][]{bos07}, presenting orbital changes in morphology \citep{rib08,mol12} and flux  \citep[see, e.g.,][]{bos05,tak09}, respectively. The source is a compact binary with period $P\approx 3.91$~d, semi-major
axis $a\approx 2.1\times 10^{12}$~cm, and eccentricity $e\approx 0.35$, and hosts a O6.5V star and an
undetermined compact object (CO) of a few solar masses \citep{cas05}. The source was initially thought to be
a radio-loud X-ray binary \citep{mar98}, and a gamma-ray emitting microquasar later on \citep{par00}. The
proposed microquasar nature and gamma-ray association led to the development of microquasar jet models for
the non-thermal emitter \citep[e.g.,][]{par00,bos04,par06,bed06,der06,kha08}, although  the lack of
accretion signatures in X-rays, among other source features, and the gamma-ray detection of
PSR~B1259$-$63/LS2883 \citep[][a similar source hosting a non-accreting pulsar]{aha05b}, led to the suggestion that LS~5039
hosted a young non-accreting pulsar \citep{mar05,dub06}. 

Orbital radial-velocity measurements have not
yet determined the LS~5039 CO mass \citep{cas05,sar11}. Several works have studied the structure of its
non-thermal emitter using modeling and/or radio \citep{rib08,bos09,mol12}, ultraviolet \citep{szo12}, X-ray
\citep{bos07,bos10,szo11,zab11}, and gamma-ray data \citep[e.g.,][]{bos08,kha08,cer10,zab13,dub15}, as well
as hydrodynamical simulations \citep{per10,bos15}. These studies show, for instance, that the emitter is likely
extended and/or relatively far from the CO, but they could not clearly favor any particular scenario.

Recently, \cite{yon20} have presented evidence of X-ray pulsations in LS~5039. Despite the relatively low
statistical significance of the detection, and an unexpectedly long period of $\approx 9$~s, these results
add significant weight to the pulsar scenario. As the pulsation period is rather long, \cite{yon20} 
suggest that the neutron star is in fact a magnetar, although the very young age estimated for the
CO, $\sim 500$~yr, seems to be at odds with the lack of evidence of the presence of a supernova remnant (SNR)
\citep{mol12b}. We note that finding a much shorter period was not possible for \cite{yon20} as the data
statistics only allowed the search of periods $>1$~s.

Regardless of the specific pulsar nature, if a pulsar wind transports the energy from the CO to the non-thermal
emitter, the wind can radiate much of its energy even before being shocked by the stellar wind. In particular, inverse Compton (IC)
scattering off stellar photons by an ultrarelativistic cold pulsar $e^\pm$-wind \citep{bog00} is very efficient in a compact binary
\citep[see, e.g.,][]{bal00,sie05,kha07,sie07,cer08,kha11,hu20}, and in LS~5039 the expected IC fluxes may be higher than the observed ones
\citep{cer08}. The total non-thermal luminosity of LS~5039 is consistently high all along the orbit, at a level $L_{\rm NT}\approx
10^{36}$~erg~s$^{-1}$ (mostly released in the MeV-GeV range, \citealt{fer09,col14}; at 2.1~kpc, \citealt{gai18,lur18}), which is
hardly compatible with Doppler boosting (Db) models (see below), meaning that the pulsar wind luminosity $L_{\rm p}$ must indeed be $>L_{\rm NT}$. 

In this work we revisit the (weakly magnetized) cold pulsar $e^\pm$-wind model in the context of LS~5039 to constrain the hypothetical wind
properties, using for the first time the most recent data in X-rays and soft and hard gamma rays. We model how such a wind transports
the energy from the pulsar to its termination shock, which is produced by the interaction with the stellar wind. Unlike previous
works, to compute the geometry of the stellar-pulsar wind interaction region we include the pulsar wind momentum-flux losses and
orbital motion. As in \cite{cer08}, we consider both isotropic and anisotropic wind models, and assume that the $e^\pm$-wind does
not heat up due to IC braking. In Sects.~\ref{mod} and \ref{res} we present our wind model and its results, respectively, adopting parameter
values that minimize the $e^\pm$-wind IC emission at periastron, the orbital phase when this emission should be the brightest (i.e., the most constraining
choice). We conclude and discuss the results in Sect.~\ref{dis}. 

\section{Model}\label{mod}

The cold pulsar $e^\pm$-wind propagation, radiation, and termination location were computed including the orbit effect on
the geometry of the contact discontinuity (CD) between the pulsar and the stellar wind. A description of the model and the approach adopted for these calculations are presented in what follows. 

To derive the most stringent
constraints on the wind radiation towards the observer, we focused on periastron, when this radiation is expected to be highest  (see \citealt{cer08}; LS~5039 periastron orbital phase, $\phi=0$, is close to superior conjunction of the CO, $\phi\approx
0.058$, see \citealt{cas05}). At the same time, we were interested in a parameter choice predicting the minimum possible IC emission for that orbital phase. Thus, regarding the inclination, a value of $i\approx 45^\circ$ was chosen, which also avoids black hole masses for the CO or stellar eclipse \citep{cas05}.
Moreover, we fixed $L_{\rm p}=2\times 10^{36}$~erg~s$^{-1}$, which is $L_{\rm p}\approx
2L_{\rm NT}$; in other words, the non-thermal emitter must emit $\approx 50$\% of the available power. As $L_{\rm NT}\sim
10^{36}$~erg~s$^{-1}$ all through the orbit, Db cannot be behind the high (but then apparent) $L_{\rm NT}$ unless the emitting flow is
always roughly pointing to the observer, which is implausible in a colliding wind system (although Db can still significantly
affect the lightcurve; see, e.g., \citealt{zab13,dub15,mol20}). For comparison, the luminosities injected in the 
emitting $e^\pm$ of the shocked wind in \cite{zab13}, \cite{dub15}, and \cite{mol20}, which must be $<L_{\rm p}$, were $L_{{\rm rel-}e}=4\times
10^{35}$, $10^{35}$, and $6\times 10^{35}$~erg~s$^{-1}$, respectively. All three works accounted for Db in the shocked-wind emitter, and the last two severely underpredicted the flux in MeV (not considered by \citealt{zab13}).

We computed for all directions the energy evolution of the wind $e^\pm$, which moved from the CO along straight trajectories as
they IC scattered stellar photons. We used an iterative scheme to follow the $e^\pm$ energy, with IC energy losses and emission
determined by IC interactions in the anisotropic blackbody stellar field \citep{kha14}. Descriptions of how to calculate the
$e^\pm$ propagation, energy loss, and emission can be found in \cite{cer08} and \cite{kha11}, among others. For simplicity, a point-like star
was considered when computing the IC emission in different directions, which is reasonable as $i$ is such that gamma rays are not
eclipsed. When computing the anisotropic IC losses for all directions the interaction angle was set to be at least the angle
subtended by the star and at most its supplementary, as seen from the emitting point, accounting in this way for the actual star finite size.  

The wind magnetic field ($B$) was taken to be irrelevant both dynamically and, as a cold $e^\pm$ wind does not
produce synchrotron radiation, radiatively. The explored initial Lorentz factors of the wind were $\Gamma_{\rm p,0}\in 10-10^5$. Gamma-ray
absorption and reprocessing were not included in the calculations because $\Gamma_{\rm p,0}\gg 10^5$ is needed for efficient
pair creation and cascading in the stellar photon field. Not  accounting for such a high $\Gamma_{\rm p,0}$ is reasonable as in LS~5039 the reprocessed gamma-ray luminosity of an
$e^\pm$-wind should be still a significant fraction of $L_{\rm p}$, peaking around multi-GeV energies
\citep[see][for a similar system]{sie05}, which is at odds with the rather low flux of LS~5039 at $\sim 10$~GeV \cite[see the
overall gamma-ray spectrum in, e.g.,][]{had12,col14}. Not   accounting for gamma-ray reprocessing simplifies the calculations
greatly as otherwise one would need a detailed knowledge of the background medium \citep[for these process complexities
in LS~5049, see, e.g.,][]{aha06b,dub06b,bed06,sie07,kha08,bos08,bos08b,cer10,bos11}.

The IC braking effect on $\Gamma_{\rm p}$ was computed for all the possible trajectories up to three times
the orbital separation (i.e., $3\times R_{\rm orb}$), and the shape of the two-wind CD was calculated including
the orbit effect. Five approximations were adopted for the initial $e^\pm$-energy and for the energy
and momentum flux angular dependence: (case A) an isotropic $e^\pm$-wind; (cases B and B') a phenomenological
axisymmetric $e^\pm$-wind with constant $\Gamma_{\rm p,0}$ and 
\begin{equation}
{\rm d}L_{\rm p}/{\rm d}\Omega\propto\sin^2\theta_{\rm p}
\end{equation}
for B and
\begin{equation}
{\rm d}L_{\rm p}/{\rm d}\Omega\propto\sin^4\theta_{\rm p} 
\end{equation}
for B', being $\theta_{\rm p}$ the angle with the pulsar wind symmetry
axis; and (cases C and C') a more physical axisymmetric wind with 
\begin{equation}
{\rm d}L_{\rm p}/{\rm d}\Omega\propto\Gamma_{\rm
p,0}(\theta_{\rm p})=(10+\bar\Gamma_{\rm p,0}\sin^2\theta_{\rm p})  
\end{equation}
for C and 
\begin{equation}
{\rm d}L_{\rm p}/{\rm d}\Omega\propto\Gamma_{\rm
p,0}(\theta_{\rm p})=(10+\bar\Gamma_{\rm p,0}\sin^4\theta_{\rm p})
\end{equation}
for C' (see, e.g., \citealt{cer08,tch16}, and
\citealt{bog02} in the context of Crab). The explored values for $\bar\Gamma_{\rm p,0}$ were $\in 10-10^5$ (as they were for
$\Gamma_{\rm p,0}$). The line-of-sight angle of the $e^\pm$-wind axis in cases B and C was set to $\psi=20^\circ$, and in
cases B' and C' to $\psi=40^\circ$, not to violate the observed fluxes. Several $e^\pm$-wind orientations
were tested fulfilling $\psi\approx 20^\circ$ and $40^\circ$, and to be conservative we chose those with the lowest IC fluxes. The  probability
of having a smaller $\psi$-value was $P\lesssim 6$\% for cases B and C, and $P\lesssim 24$\% for cases B' and C'. The stellar wind was assumed to be spherically symmetric, although a
prescription for its momentum flux $\propto\sin^2\theta_{\rm w}$ was also tested, with the plane defined by $\theta_{\rm
w}=\pi/2$ passing through the pulsar center (the most conservative case); both prescriptions yielded very similar results. 

The stellar wind was assumed to be cold, with momentum flux of 
\begin{equation}
\dot{\vec{\rm p}}_{\rm w}=\rho_{\rm w}v_{\rm
w}\vec{\rm v}_{\rm w}\,, 
\end{equation}
and velocity of 
\begin{equation}
\vec{\rm v}_{\rm w}=v_\infty(1-R_*/d)^{\beta}\hat{\bf
r}-\Omega\,d\sin\theta\hat{\vec{\phi}} 
\end{equation}
in the non-inertial frame rotating with the orbit angular velocity ($\Omega$),
where $\theta$ is the angle to the orbit normal, $R_*$ the stellar radius, and $d$ the stellar distance. 
To normalize $\dot{\vec{\rm p}}_{\rm w}$, the maximum possible mass-loss rate for
this wind, $\dot{M}=7\times 10^{-7}$~M$_\odot$~yr$^{-1}$, was adopted \citep{cas05}, minimizing the distance
covered by the unshocked $e^\pm$-wind, and thus its IC emission. Nevertheless, the $\dot{M}$-dependence of the $e^\pm$-wind IC luminosity is weak, as the pulsar wind facing the observer at periastron typically loses most of its energy (see below).
The stellar wind velocity at infinity $v_\infty$ was fixed to $2.44\times
10^8$~cm~s~$^{-1}$, and $\beta$ to 1, while the minimum radial velocity was set to the escape velocity,
$\approx 10^8$~cm~s~$^{-1}$ \citep{mcs04}. 

The $e^\pm$-wind, little affected by Coriolis forces due to its
high speed, was assumed with zero pressure, radial from the pulsar, and with momentum flux of 
\begin{equation}
\dot{\vec{\rm p}}_{\rm p}=\rho_{\rm
p}\Gamma^2_{\rm p}v^2_{\rm p}\hat{\bf r}_{\rm p}\,, 
\end{equation}
and velocity of 
\begin{equation}
\vec{\rm v}_{\rm
p}=(1-1/\Gamma^2_{\rm p})^{1/2}c\,.
\end{equation}
The $e^\pm$-wind anisotropy was introduced via $\rho_{\rm
p}$ (B and B') and $\Gamma_{\rm p,0}$ (C and C'). Losses affected $\Gamma_{\rm p}$.

The CD was computed using an approach based on the thin shell axisymmetric approximation (see, e.g.,
\citealt{ant04}), but modified to account
for the lack of axisymmetry due to the orbit-related Coriolis force and the wind anisotropy. 

First, the CD
stagnation point (SP), at which 
\begin{equation}
\dot{\vec{\rm p}}_{\rm p}+\dot{\vec{\rm p}}_{\rm w}=0\,, 
\end{equation}
is  found. We
looked for the SP assuming that it was on the orbital plane, in a direction from the pulsar tilted from the
pulsar-star direction clockwise by an angle 
\begin{equation}
\tau\approx \frac{\omega_{\rm orb}R_{\rm
orb}(1+\eta^{1/2})}{|\vec{\rm{v}}_{\rm w}|}\,, 
\end{equation}
where $\omega_{\rm orb}$ is the orbit angular velocity and $\eta=(L_{\rm
p}/\dot{M}|\vec{\rm v}_{\rm w}|c)$ is the momentum rate ratio of the pulsar and the stellar wind. This
approximation works well for small $\tau$, so we adopted this method in our calculations as a first-order approximation. Once found, we took the SP position as the initial
one and computed the paths shaping the CD in all directions through an iterative process. The
iterative steps in each path provide   segments that  characterize the CD surface step by step. To find the directions of the new segments, we used the relations
\begin{equation}
|\dot{\vec{\rm p}}_{\rm
w}|\sin^2\alpha_{\rm w}=|\dot{\vec{\rm p}}_{\rm p}|\sin^2\alpha_{\rm p}
\label{rel1} 
\end{equation}
and 
\begin{equation}
\dot{\vec{\rm p}}_{\rm
s}=\dot{\vec{\rm p}}_{\rm w}\sin\alpha_{\rm w}+\dot{\vec{\rm p}}_{\rm p}\sin\alpha_{\rm p} 
\label{rel2}
\end{equation}
at each step, where $\alpha_{\rm w}$ and $\alpha_{\rm p}$ are the complementary angles between the stellar and the pulsar wind
directions and the CD normal, respectively, and $\dot{\vec{\rm p}}_{\rm s}$ is the momentum
flux remaining after wind collision, at the segment location on the CD. The thin shell approximation allows the cancellation of the opposing momentum components. Once a
direction on the CD surface from the SP is chosen, the next point is defined by the local
direction of $\dot{\vec{\rm p}}_{\rm s}$ and a sufficiently small distance in that direction. Even if $\alpha_{\rm w}$ and $\alpha_{\rm p}$ are not known, the direction of $\dot{\vec{\rm p}}_{\rm s}$ can be computed by
eliminating one of the $\alpha$ parameters using Eq.~(\ref{rel1}), and renormalizing $\dot{\vec{\rm p}}_{\rm s}$ to
eliminate the other one from Eq.~(\ref{rel2}).

To account for the shocked region of the $e^\pm$-wind between the termination shock and the CD, the $e^\pm$-wind was
terminated at $\approx 2/3$ of the distance from the pulsar to the CD (see \citealt{bog08} for cases with $\eta\lesssim
0.01$, although if $B$ were strong it would largely modify the overall interaction structure, see \citealt{bog19}). In the presence of orbital motion, analytical and numerical calculations show that a strong shock should terminate the pulsar wind due
to Coriolis forces in the directions away from the star \citep[see, e.g.,][]{bos11b,bos12,bos15,hub20}. Therefore, the unshocked
pulsar wind, and the CD itself, were simply assumed to stop in these directions at the distance where such a {Coriolis shock} is
expected to form. Since we focus on periastron, our results are not affected by this assumption as the relevant line
of sight does cross the computed CD at that phase.

\section{Results}\label{res}

The cold pulsar $e^\pm$-wind IC fluxes were computed for cases A, B and B', and C and C'. The pulsar orientation in B and C, and in B' and C', was
chosen such that $\psi\approx20^\circ$ ($P\lesssim 6$\%) and $40^\circ$ ($P\lesssim 24$\%), respectively. The remaining parameter values were as follows: $R_*=10\,R_\odot$; effective stellar
temperature $T=4\times 10^4$~K; $e=0.35$; $P=3.91$~d; $\phi=0$ ($R_{\rm orb}\approx 1.4\times 10^{12}$~cm); $i=45^\circ$;
$\dot{M}=7\times 10^{-7}$~M$_\odot$~yr$^{-1}$; $v_\infty=2.44\times 10^8$~cm~s~$^{-1}$ and $\beta=1$; $L_{\rm p}=2\times
10^{36}$~erg~s$^{-1}$; and $\Gamma_{\rm p,0}\in 10-10^5$. 

\begin{figure}        
\includegraphics[width=7.8cm]{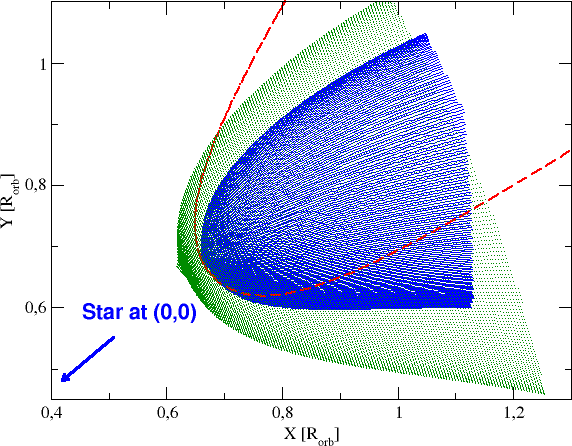}
\includegraphics[width=9.2cm]{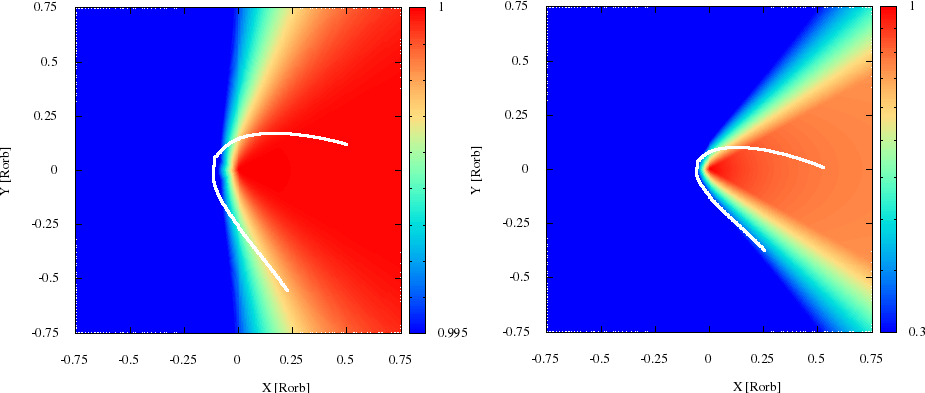}
\caption{{\bf Top panel:} Orbital plane projection of the computed paths on the CD surface, at periastron, for case A and $\Gamma_{\rm p,0}=10$ (green; more extended) and $10^5$ (blue; more compact), including orbital motion. The red dashed line shows the profile of the structure for $\Gamma_{\rm p,0}=10^5$ when orbital motion is not included. The projected line of sight points along $-\hat{y}$ and the star is at $(0,0)$.
{\bf Bottom panel:} Color maps of the orbital plane of the remaining energy fraction, after IC losses, for particles propagating from the pulsar in all directions, for $\Gamma_{\rm p,0}=10$ (left) and $10^5$ (right). The white solid lines are the profiles of the projected CD on the orbital plane at periastron for case A including orbital motion. The pulsar is at $(0,0)$. {\bf Both panels:} $x$- and $y$-axis units are the orbital distance.}
\label{map}
\end{figure}

The paths characterizing the CD surface for case A are shown in the top panel of  Fig.~\ref{map}   for $\Gamma_{\rm
p,0}=10$ and $10^5$, with and without orbital motion (the CD shapes for B and C, and B' and C' were very similar, although a detailed
discussion is beyond the scope of this work). The CD shrinks by a factor of $\sim 2$ due to stronger IC losses for
$\Gamma_{\rm p,0}=10^5$, as the cooling time is $\propto \Gamma_{\rm p,0}^{-1}$ in the explored range. Orbital motion clearly affects the
CD geometry on the scales of the orbital separation for the parameter values adopted, although the computed $e^\pm$-wind IC
fluxes without orbital motion (see the dashed line in the top panel of Fig.~\ref{map}) were almost the same. The bottom panel of Figure~\ref{map}  
shows two maps of the fraction of the energy remaining in the $e^\pm$ after IC losses for case A, with $\Gamma_{\rm
p,0}=10$ (right) and $10^5$ (left), together with the corresponding orbital-plane CD shapes. As expected,
for $\Gamma_{\rm p,0}=10$ $e^\pm$ lose $\sim 1$\% of their energy or less when reaching the CD, whereas for $\Gamma_{\rm
p,0}=10^5$ they lose $\sim 80$\% towards the star, and $\sim 20$\% away from it. 

\begin{figure}        
\includegraphics[width=8.8cm]{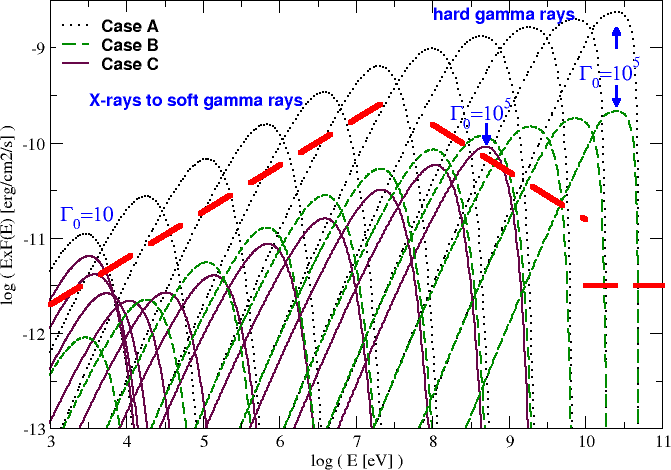}
\caption{Computed spectral energy distributions of the IC cold pulsar wind emission, at periastron, for 11 different 
$\Gamma_{\rm p,0}$ values from $\Gamma_{\rm p,0}=10$ to $10^5$, for cases A (dotted lines), B (dashed lines), and C (solid lines) (cases B' and C' are very similar; see main text). A schematic spectral energy distribution based on the  observed X-ray to gamma-ray emission around superior conjunction (adapted from fig.~9 in \citealt{col14}; see references therein), which is close to periastron, is also shown.}
\label{sed}
\end{figure}

The $e^\pm$-wind spectral energy distributions are shown in Fig.~\ref{sed} for $\Gamma_{\rm p,0}\in 10-10^5$, and for cases A, B, and C. The spectral energy distributions for cases B' and C' (not shown) are very similar to those of B and C, as expected given that the $\psi$-value choice is determined by the same observational constraints.  
An
isotropic $e^\pm$-wind (A) overpredicts the observed fluxes by at least a factor of $\approx 3$. The anisotropic $e^\pm$-wind of cases B and C, with $\psi\lesssim 20^\circ$
between the wind symmetry axis and the line of sight, does not overpredict the observed X-ray or $\gtrsim 100$~MeV fluxes if $\Gamma_{\rm p,0}\notin 10^2-10^3$. The IC flux of cases B and C is also lower than the
observed $\sim 0.1-30$~MeV fluxes by a factor of $\approx 3-5$ for $\Gamma_{\rm p,0}\in 10^2-10^3$, which allows the relaxation of the constraint on $\psi$ from $\lesssim 20^\circ$ to $\lesssim 45^\circ$ ($P\lesssim 30$\%) for $L_{\rm p}\gtrsim 2\times 10^{36}$~erg~s$^{-1}$. In the anisotropic $e^\pm$-wind model of cases B' and C', the observed X-ray and $\gtrsim 100$~MeV fluxes constrain $\psi$ to $\lesssim 40^\circ$ ($P\lesssim 24$\%) if $\Gamma_{\rm p,0}\notin 10^2-10^3$. On the other hand, the observed $\sim 0.1-30$~MeV fluxes do not constrain $\psi$ at all in cases B' and C' for $\Gamma_{\rm p,0}\in 10^2-10^3$ if $L_{\rm p}\gtrsim 2\times 10^{36}$~erg~s$^{-1}$.
The X-ray data, having high statistics, constrains a narrow spectral component to a level of a $\sim 10$\% of
the observed fluxes \citep[see, e.g.,][]{bos07}. Thus, Cases B and B' are also constrained by X-ray data even though the predicted flux is $\sim 3$ times lower than observed.

\section{Conclusions and discussion}\label{dis}

\subsection{Weakly magnetized cold wind model}

If LS~5039 has a weakly magnetized cold pulsar $e^\pm$-wind carrying the energy to the non-thermal emitter, this wind
cannot be isotropic as this is inconsistent with observations, even for ratios of the total broadband non-thermal radiation luminosity to the pulsar wind power of 
$L_{\rm NT}/L_{\rm p}\sim 1$. 

An anisotropic wind with angular dependence $\propto\sin^2\theta_{\rm p}$ does not violate the observational constraints, but requires either a very high $L_{\rm
NT}/L_{\rm p}\sim 1$, even for the most favorable $\Gamma_{\rm p,0}$ values, or an unlikely orientation. For cases $B$ and $C$ and $\Gamma_{\rm p,0}\notin 10^2-10^3$,
even for an unrealistic $L_{\rm NT}/L_{\rm p}\approx 0.5$ a rather improbable $\psi\lesssim 20^\circ$ ($P\lesssim 6$\%) is required for the predictions to be
consistent with X-ray and gamma-ray data. For the same cases but where $\Gamma_{\rm p,0}\in 10^2-10^3$, $\psi$ is less constrained, $\lesssim 45^\circ$ ($P\lesssim 30$\%),
if  $L_{\rm NT}/L_{\rm p}\approx 0.5$ is kept. Smaller $\psi$-values relax the energetic constraints when $\Gamma_{\rm p,0}\in 10^2-10^3$, but reduce the associated  probability: for $L_{\rm NT}/L_{\rm p}\approx 0.1$, $\psi\lesssim
20^\circ$ and thus $P\lesssim 6$\%. We note that $L_{\rm NT}/L_{\rm p}\approx 0.1$ is still quite high,  typically values of $L_{\rm{rel-}e}/L_{\rm p}\sim 0.1$ are adopted in the
modeling of the whole LS~5039 non-thermal emission, resulting in an even lower $L_{\rm NT}/L_{\rm p}$ value. We recall that invoking Db can reduce the energetic requirements in some orbital phases, but not everywhere in the orbit.

If the wind is even more anisotropic, with angular dependence $\propto\sin^4\theta_{\rm p}$, the situation becomes less dramatic,
although quite high $L_{\rm NT}/L_{\rm p}$ values are still needed in general. For cases $B'$ and $C'$ and $\Gamma_{\rm p,0}\notin
10^2-10^3$, setting $L_{\rm NT}/L_{\rm p}\approx 0.5$ implies that $\psi\lesssim 40^\circ$ ($P\lesssim 24$\%) does not   violate the observed
fluxes, whereas for $\Gamma_{\rm p,0}\in 10^2-10^3$ the angle $\psi$ is not constrained at all. Adopting $\psi$-values significantly smaller
than $\approx 40^\circ$, $L_{\rm NT}/L_{\rm p}$ becomes virtually unconstrained for $\Gamma_{\rm p,0}\in 10-10^5$ in B', and for
$\Gamma_{\rm p,0}\in 10^2-10^5$ in case C' (in C' taking $\Gamma_{\rm p,0}\lesssim 10^2$ still overpredicts the X-ray fluxes). Therefore, in cases
B' and C' the data are less restrictive than B and C of the $e^\pm$-wind properties, but some fine-tuning in $\psi$ or $\Gamma_{\rm p,0}$ is still
needed if a high $L_{\rm NT}/L_{\rm p}$ is to be avoided.

Despite all this discussion on wind anisotropy, it is worth noting that fine-tuning the wind orientation is less effective in
relaxing the energetic requirements on $L_{\rm p}$ if the pulsar precesses \citep{sta00}. In addition, the wind anisotropy is likely
to be more complex than just described \citep[see, e.g.,][]{phi18}, which can smear its effects and thus make wind orientation fine-tuning less effective in relaxing the observational constraints.

To explore the $e^\pm$-multiplicity $\kappa$ in the studied wind model, one can take for instance $\Gamma_{\rm p,0}=3\times 10^2$
and $L_{\rm p}\sim 10^{37}$~erg~s$^{-1}$, values allowed by observations for $\psi\lesssim 20^\circ$ in B and C, and in a broader range of 
$\psi$-values in B' and C'. With this choice of parameters, $\kappa$ should be $\sim 10^6$ ($\sim 10^9$) for a 10~km radius pulsar
with $10^{12}$~G ($10^{15}$~G) surface $B$ and $10^{-2}$~s (10~s) period.  

A weakly magnetized cold baryonic wind is an alternative to a pure $e^\pm$-wind model. Such a wind would require that baryons
reach $\Gamma_{\rm p,0}\gtrsim 5\times 10^4$ ($\gtrsim 5\times 10^7$ -10~s-) for $L_{\rm p}\gtrsim 2\times 10^{36}$~erg~s$^{-1}$, while
$\kappa\lesssim 10$ ($\lesssim 10^4$ -10~s-), so as not to violate the $\gtrsim 10$~GeV ($\gtrsim 1$~TeV -10~s-) fluxes for an isotropic wind, and
$\lesssim 10^2$ ($\lesssim 10^5$ -10~s-) for an anisotropic wind with $\psi\approx 20^\circ$ (B and C) or $\approx 40^\circ$ (B' and C').
We note that this scenario requires very efficient proton-to-$e^\pm$ transfer at wind termination, as multiplicity limits become
tighter for higher proton wind luminosity budgets. We consider here protons solely for energy transport as a
non-thermal proton emitter in LS~5039 is   likely to be radiatively inefficient due to not enough ambient photon energy, and
radiation and wind densities \citep{bos09b}. 

\subsection{Magnetized winds}

A strongly magnetized flow produced by a young magnetar in LS~5039 was proposed by \cite{yon20} prompted by the evidence of a $\approx 9$~s period in X-ray data. 
As indicated in Sect.~\ref{intro}, the young magnetar hypothesis put forward in that work is at odds with the lack of evidence for a SNR, and the physics of such a
flow would be more uncertain than that of more standard pulsar wind models. Interestingly, it is also possibile to  envision an exotic explanation for the $\approx 9$~s
period unrelated to pulsar rotation.  LS~5039 might instead  host  not one CO, but an extremely close CO binary of semi-major axis $\approx 10^9$~cm. Such a
binary, given the mass function of the system \citep{cas05}, would most likely be formed by two neutron stars. The neutron stars might have rotation periods much
shorter than $\approx 9$~s, for instance in the millisecond range. Now, whether such a configuration is plausible from the point of view of stellar evolution is to be studied elsewhere \citep[see, e.g.,][in the context of gravitational wave detections of related systems]{abb20}.

Nevertheless, for   the single-CO scenario a pulsar wind with a dynamically significant magnetic field would be in agreement with recent research on pulsar winds, which proposes average magnetization parameters $\sigma\sim 1$ instead of $\sim 10^{-2}-10^{-3}$ \citep{ama20}, where
\begin{equation}
\sigma=\frac{L_{\rm p}}{\Gamma_{\rm p,0}\dot{m}c^2}\,, 
\end{equation}
with $\dot{m}$ being the wind mass rate. 
Unfortunately, the complexity of a magnetized wind \citep[as exemplified for instance
in the simulations by][]{phi18} does not allow the use of a simple, but still physically consistent, prescription for the wind like the one employed here for the cold, matter-dominated case. However, we can already show that a strongly magnetized wind may be, given the observational constraints, the most favored scenario in LS~5039: 

As in the case of a cold baryonic wind, a magnetized wind would not produce prominent narrow X-ray and gamma-ray spectral features if
the wind $e^\pm$ did not get energized by $B$-dissipation until wind termination (and $e^\pm$-energy redistribution). However, for a strongly magnetized wind, the
constraints on $\kappa$ and $\Gamma_{\rm p,0}$ are looser than in the baryonic wind case, as now $\Gamma_{\rm p,0}$ is also not constrained. For $\Gamma_{\rm p,0}\lesssim 10^4$, we obtain $\kappa\Gamma_{\rm p,0}\lesssim 10^7$ ($\lesssim 10^{10}$ -10~s-) for case A,
implying $\sigma\gtrsim 10$. For $\Gamma_{\rm p,0}\gtrsim 10^4$, $\sigma$ must be higher due to tighter observational constraints above the GeV range. 
For cases B, C, B', and C' the observational constraints on a weakly magnetized wind are looser for specific orientations, and thus
less demanding on sigma, but high sigma values render wind orientation fine-tuning (which  has its caveats, as already mentioned)
unnecessary. Therefore, we conclude that a high-sigma wind is less constrained and thus seems more favorable than the other two scenarios. 

\subsection{Hot winds}

It is possible that, relatively early in its propagation, the pulsar wind turns  a significant amount of energy (magnetic or kinetic) into non-thermal energy,  meaning that the
non-thermal emitter would {start} much closer to the pulsar than the wind termination shock \citep[see, e.g.,][for the presence of non-thermal particles in the
wind of LS~5039 due, e.g., to magnetic dissipation, IC cascades]{sie07,pet11,der12}. This scenario is perhaps the hardest to model, and may not be
consistent with evidence of fast $e^\pm$-wind acceleration in the Crab pulsar \citep[see][]{aha12}, although in a high-mass binary system the unshocked pulsar wind
has a very different environment than in an isolated pulsar. On the other hand, it is worth noting that observations of LS~5039 strongly suggest that significant particle
acceleration is still required far from the CO as the X-ray and particularly the very high-energy emission seem to originate in the binary outskirts (see
Sect.~\ref{intro}). This peripheric accelerator must be very efficient as acceleration rates close to the electrodynamical limit have been inferred from the
very high-energy data \citep[see][]{kha08}. Nevertheless, it cannot be discounted that the bulk of the $\sim 0.1-30$~MeV emission may still be produced in a {hot} wind, relatively close to the pulsar \citep[as proposed in][effectively similar to the cold $e^\mp$-wind case with $\Gamma_{\rm p,0}\in 10^2-10^3$ discussed
here]{der12}\footnote{The close pulsar binary sketched above could generate a {hot} two-pulsar combined wind.}.

\subsection{Similar sources}

We finish by noting that, in addition to LS~5039, four other high-mass gamma-ray binaries, LS~I~+61~303,
1FGL~J1018.6$-$5856, LMC~P3, and 4FGL~J1405$-$6119, are approximately as compact and at least as powerful as
LS~5039, with a relatively similar phenomenology and unknown CO \citep[see Sect.~1 in][and references
therein]{mol20}. Although all these sources deserve specific studies of their own, an approach such as the one presented here can be
helpful to constrain the properties of a hypothetical pulsar wind powering their non-thermal emitter. 

\section*{Acknowledgements}
We thank the anonymous referee for constructive and useful comments that helped to improve the manuscript.
We are grateful to Dmitry Khangulyan for insightful comments on this work.
V.B-R. acknowledges financial support by the Spanish Ministerio de Econom\'ia, Industria y Competitividad (MINEICO/FEDER, UE) under grant AYA2016-76012-C3-1-P, from the State Agency for Research of the Spanish Ministry of Science and Innovation under grant PID2019-105510GB-C31 and through the ''Unit of Excellence Mar\'ia de Maeztu 2020-2023'' award to the Institute of Cosmos Sciences (CEX2019-000918-M), and by the Catalan DEC grant 2017 SGR 643. V.B-R. is Correspondent Researcher of CONICET, Argentina, at the IAR.

\bibliographystyle{aa}
\bibliography{biblio}

%\label{lastpage}

\end{document}